\documentclass{article}
\usepackage[T1]{fontenc}
\usepackage[latin9]{inputenc}
\usepackage{amsmath}
\usepackage{amssymb}
\usepackage{graphicx}
\usepackage{comment}

\usepackage{graphicx,bm} 
\usepackage{subfigure} 

\usepackage{natbib}

\usepackage{algorithm}
\usepackage{algorithmic}

\usepackage{hyperref}

\makeatletter

\providecommand{\tabularnewline}{\\}

\usepackage[accepted]{icml2013}

\newtheorem{theorem}{Theorem}

\newtheorem{lemma}{Lemma}
\newtheorem{definition}{Definition}

\icmltitlerunning{Spectral Compressed Sensing via Structured Matrix Completion}

\begin{document}
\twocolumn[

\icmltitle{Spectral Compressed Sensing via Structured Matrix Completion}

\icmlauthor{Yuxin Chen}{yxchen@stanford.edu}
\icmladdress{Department of Electrical Engineering, Stanford University, Stanford, CA 94305, USA}
\icmlauthor{Yuejie Chi}{chi@ece.osu.edu}
\icmladdress{Electrical and Computer Engineering, The Ohio State University, Columbus, OH 43210, USA }

\icmlkeywords{compressed sensing, spectral estimation, matrix completion, Hankel matrix, super-resolution, harmonic retrieval}

\vskip 0.3in
]

\begin{abstract}
The paper studies the problem of recovering a spectrally sparse object from a small number of time domain samples. Specifically, the object of interest with ambient dimension $n$ is assumed to be a mixture of $r$ complex multi-dimensional sinusoids, while the underlying frequencies can assume any value in the unit disk. Conventional compressed sensing paradigms suffer from the {\em basis mismatch} issue when imposing a discrete dictionary on the Fourier representation. To address this problem, we develop a novel nonparametric algorithm, called enhanced matrix completion (EMaC), based on structured matrix completion. The algorithm
starts by arranging the data into a low-rank enhanced form with
multi-fold Hankel structure, then attempts recovery via nuclear norm minimization. Under mild incoherence conditions, EMaC allows perfect recovery as soon as the number of samples exceeds the order of $\mathcal{O}(r\log^{2} n)$. We also show that, in many instances, accurate completion of a low-rank multi-fold Hankel matrix is possible
when the number of observed entries is proportional to the information theoretical
limits (except for a logarithmic gap). The robustness of EMaC against bounded noise and its applicability to super resolution are further demonstrated by numerical experiments.
\end{abstract}

\section{Introduction}

A large class of practical applications features high-dimensional objects that can be modeled or  approximated by a superposition of spikes in the spectral (resp. time) domain, and involves estimation of the object from its time (resp. frequency) domain samples.  A partial list includes medical imaging \cite{Lustig2007sparse}, radar systems \cite{potter2010sparsity}, seismic imaging \cite{Borcea2002imaging}, microscopy \cite{schermelleh2010guide}, etc. The data acquisition devices, however, are often limited by physical and hardware constraints, precluding sampling with the desired resolution. It is thus of paramount interest to reduce the sensing complexity while retaining recovery resolution. 

Fortunately, in many instances, it is possible to recover an object even when the number of samples is far below the ambient dimension, provided that the object has a parsimonious representation in the transform domain. In particular, recent advances in compressed sensing (CS) \cite{CandRomTao06} popularize the nonparametric methods based on convex surrogates. Such tractable methods do not require prior information on the model order, and are often robust against noise.

Nevertheless, the success of CS relies on sparse representation or approximation of the object of interest in a finite discrete dictionary, while the true parameters in many application are actually specified in a {\em continuous} dictionary. For concreteness,  consider an object $x\left(\boldsymbol{t}\right)$ that is a weighted sum of $K$-dimensional sinusoids at $r$ distinct frequencies $\{\boldsymbol{f}_{i}\in [0,1]^K:1\leq i \leq r\}$. Conventional CS paradigms operate under the assumptions that these frequencies lie on a pre-determined grid on the unit disk. However, cautions need to be taken when imposing a discrete dictionary on continuous frequencies, since nature never poses the frequencies on the pre-determined grid, no matter how fine the grid is \cite{Chi2011sensitivity, duarte2012spectral}. This issue, known as \textit{basis mismatch} between the true frequencies
and the discretized grid, results in loss of sparsity due to spectral
leakage along the Dirichlet kernel, and hence degeneration in the performance
of CS algorithms. While one might impose finer gridding to mitigate this weakness, this approach often leads to numerical instability and high correlation between dictionary elements, which significantly weakens the advantage of these CS approaches \cite{TangBhaskarShahRecht2012}. 

In this paper, we explore the above \emph{spectral compressed sensing} problem, which
aims to recover a spectrally sparse object from a small set of time-domain samples. The underlying (possibly {\em multi-dimensional}) frequencies can assume any value in the unit disk, and need to be recovered with infinite precision. To address this problem, we develop a {\em nonparametric} algorithm, called enhanced matrix
completion (EMaC), based on structured matrix completion. Specifically, EMaC starts by converting the data samples
into an enhanced matrix with ($K$-fold) Hankel structures, and then solves a nuclear-norm minimization program to complete the enhanced matrix. We show that, under mild incoherence conditions, EMaC admits exact recovery from $\mathcal{O}(r\log^{2}n)$ random samples, where $r$ and $n$ denote respectively the spectral sparsity and the ambient dimension. Additionally, we provide theoretical guarantee for the low-rank Hankel matrix completion problem, which is of great importance in control, natural language processing, computer vision, etc. To the best of our knowledge, our results provide the first theoretical bounds that are close to the information theoretic limit.  Furthermore, numerical experiments demonstrate that our algorithm is robust against noise and is applicable to the problem of super resolution. 

\section{Connection to Prior Art} 
The spectral compressed sensing problem is closely related to {\em harmonic retrieval}, which seeks to extract the underlying frequencies of an object from a collection of its time-domain samples. This spans many signal processing applications including radar localization systems \cite{NionSid2010}, array imaging systems \cite{Borcea2002imaging}, wireless channel sensing \cite{SayeedAazhang1999,gedalyahu2011multichannel}, etc. In fact, if the time-domain representation of an object can be estimated accurately, then its underlying frequencies can be identified using harmonic super-resolution methods.

Conventional approaches for these problems, such as ESPRIT \cite{RoyKailathESPIRIT1989} and the matrix pencil method \cite{Hua1992}, are based on the eigenvalue
decomposition of covariance matrices constructed from {\em equi-spaced} samples, which can accommodate infinite frequency
precision. One weakness of these techniques lies in that they require prior information on the model order, that is, the
number of underlying frequency spikes or, at least, an estimate of it.
Besides, their performance largely depends on the knowledge of
noise spectra; some of them are unstable in the presence of noise and outliers \cite{DragVetterliBlu2007}.

Nonparametric algorithms based on convex optimization differ from
the above parametric techniques in that the model order does not need to be specified
\emph{a priori}. Recently, Cand\`es and Fernandez-Granda \yrcite{CandesFernandez2012SR}
proposed a total-variation minimization algorithm to super-resolve a sparse object from frequency samples at the low end of its spectrum. This algorithm allows accurate
super-resolution when the point sources are appropriately separated, and is stable against noise \cite{CandesFernandez2012SRNoisy}. Inspired by this approach, Tang et. al. \yrcite{TangBhaskarShahRecht2012}
developed an atomic norm minimization algorithm for line spectral estimation from $\mathcal{O}(r\log r\log n)$ time domain samples. This work is limited to 1-D frequency models and assumes randomness in the data model. 

In contrast, our approach can accommodate multi-dimensional frequencies, and only assumes randomness in the observation basis. The algorithm is inspired by the recent advances of matrix completion (MC) problem, which aims at
recovering a low-rank matrix from partial entries. It
has been shown \cite{ExactMC09,Gross2011recovering}
that exact recovery is possible via nuclear norm minimization, as
soon as the number of observed entries is on the order of the information
theoretic limit. Encouragingly, this line of algorithms is also robust
against noise and outliers \cite{Negahban2012restricted}.
Nevertheless, the theoretical guarantees of these algorithms do not apply to the more
structured observation models associated with Hankel structure. Consequently,
direct application of existing MC results yields pessimistic bounds
on the number of samples, which is far beyond the degrees
of freedom underlying the sparse object.

The rest of this paper is organized as follows. Section~\ref{sec:Model-and-Algorithm} describes the data model and the EMaC algorithm. Section~\ref{sec:Main-Results} presents the theoretical guarantee of EMaC, with a proof outlined in the Appendix. Numerical validation of EMaC is given in Section \ref{sec:Numerical-Experiments}. We then discuss the extension to low-rank Hankel matrix completion in Section \ref{sub:Extension-to-Hankel}, and conclude the paper in Section \ref{sec:conclusions-and-future}.

\section{Model and Algorithm\label{sec:Model-and-Algorithm}}

Assume that the object of interest $x\left(\boldsymbol{t}\right)$ can be modeled as a weighted sum of $K$-dimensional sinusoids at $r$ distinct frequencies
$\boldsymbol{f}_{i}\in[0,1]^{K}$ ($1\leq i\leq r$), i.e.
\begin{equation}
x\left(\boldsymbol{t}\right)=\sum_{i=1}^{r}d_{i}e^{j2\pi\left\langle \boldsymbol{t},\boldsymbol{f}_{i}\right\rangle}
\end{equation}
where $d_i$'s denote the complex amplitude. For concreteness, our discussion is mainly devoted to a 2-dimensional (2-D) frequency model.
This subsumes 1-D line spectral estimation as a special case, and
indicates how to address multi-dimensional models. 

\subsection{2-D Frequency Model and Problem Setup\label{sub:2-D-Frequency-Model}}
Consider a data matrix $\boldsymbol{X}=\left(x_{k,l}\right)_{0\leq k<n_1, 0\leq l<n_2}$ of size $n_{1}\times n_{2}$. Suppose each entry can be expressed as $$x_{k,l}=\sum_{i=1}^{r}d_{i}y_{i}^{k}z_{i}^{l},$$
where $y_{i}=\exp\left(j2\pi f_{1i}\right)$ and $z_{i}=\exp\left(j2\pi f_{2i}\right)$
for some set of frequency pairs $\left\{ \left(f_{1i},f_{2i}\right)\mid1\leq i\leq r\right\}$ (normalized by the Nyquist rate).
We can then write $\boldsymbol{X}$ in a matrix form as follows:
\begin{equation}
\boldsymbol{X}=\boldsymbol{Y}\boldsymbol{D}\boldsymbol{Z}^{T},\label{eq:X_MatrixForm}
\end{equation}
where $\boldsymbol{D}:=\text{diag}\left[d_{1},d_{2},\cdots,d_{r}\right]$, and $\boldsymbol{Y}$ and $\boldsymbol{Z}$ are defined as
\begin{equation}
\boldsymbol{Y}:=\left[\begin{array}{cccc}
1 & 1 & \cdots & 1\\
y_{1} & y_{2} & \cdots & y_{r}\\
\vdots & \vdots & \vdots & \vdots\\
y_{1}^{n_{1}-1} & y_{2}^{n_{1}-1} & \cdots & y_{r}^{n_{1}-1}
\end{array}\right],
\end{equation}
\begin{equation}
\boldsymbol{Z}:=\left[\begin{array}{cccc}
1 & 1 & \cdots & 1\\
z_{1} & z_{2} & \cdots & z_{r}\\
\vdots & \vdots & \vdots & \vdots\\
z_{1}^{n_{2}-1} & z_{2}^{n_{2}-1} & \cdots & z_{r}^{n_{2}-1}
\end{array}\right].\label{eq:Y_matrix_form}
\end{equation}

Suppose that there exists a location set $\Omega$ of size $m$ such
that $x_{k,l}$ is observed iff $\left(k,l\right)\in\Omega$, and assume that $\Omega$ is sampled uniformly at random.  We are interested in recovering $\boldsymbol{X}$ from its partial observation on the location set $\Omega$.

Before continuing, we introduce a few notations that will be used throughout. The spectral norm (operator norm), the Frobenius norm, and the nuclear norm (sum of singular values) of a matrix $\boldsymbol{X}$ are denoted by $\left\Vert \boldsymbol{X}\right\Vert$, $\left\Vert \boldsymbol{X}\right\Vert_{\text{F}}$ and $\left\Vert \boldsymbol{X}\right\Vert_{*}$, respectively. The inner product between two matrices is defined by $\left\langle \boldsymbol{B},\boldsymbol{C}\right\rangle =\text{trace}\left(\boldsymbol{B}^{*}\boldsymbol{C}\right)$. We let $\mathcal{P}_\Omega$ be the orthogonal projection onto the subspace of matrices that vanishes outside $\Omega$.

\subsection{Enhanced Matrix Completion (EMaC)\label{sub:Matrix-Enhancement}}

One might naturally attempt recovery by applying the matrix completion algorithms \cite{ExactMC09}
on the original data matrix, arguing that when $r$ is small,
perfect recovery of $\boldsymbol{X}$ is possible from partial measurements. Specifically, this corresponds to the following optimization
program 
\begin{align}
\underset{\boldsymbol{M}\in\mathbb{C}^{n_1\times n_2}}{\text{minimize}}\quad & \|\boldsymbol{M}\|_{*}\\
\text{subject to}\quad & \mathcal{\mathcal{P}}_{\Omega}\left(\boldsymbol{M}\right)=\mathcal{\mathcal{P}}_{\Omega}\left(\boldsymbol{X}\right),\nonumber 
\end{align}
which is the convex relaxation of the rank minimization problem. However, generic matrix completion algorithms \cite{Gross2011recovering}
require at least the order of $r\max\left(n_{1},n_{2}\right)$ samples,
which far exceeds the degrees of freedom (which is $\mathcal{O}\left(r\log n\right)$ due to a coupon collector's effect)
in our problem. What is worse, when $r>\min\left(n_{1},n_{2}\right)$
(which is possible since $r$ can be as large as $n_{1}n_{2}$),
$\boldsymbol{X}$ is no longer low-rank. This motivates us to construct
other forms (e.g. \citet{Hua1992}) that better capture the harmonic structure.

Specifically, we adopt an effective enhanced form of $\boldsymbol{X}$
based on $2$-fold Hankel structure as follows. The enhanced matrix $\boldsymbol{X}_{\text{e}}$ with respect to $\boldsymbol{X}$
is defined as a $k_{1}\times\left(n_{1}-k_{1}+1\right)$ block Hankel
matrix
\begin{equation}
\boldsymbol{X}_{\text{e}}:=\left[\begin{array}{cccc}
\boldsymbol{X}_{0} & \boldsymbol{X}_{1} & \cdots & \boldsymbol{X}_{n_{1}-k_{1}}\\
\boldsymbol{X}_{1} & \boldsymbol{X}_{2} & \cdots & \boldsymbol{X}_{n_{1}-k_{1}+1}\\
\vdots & \vdots & \vdots & \vdots\\
\boldsymbol{X}_{k_{1}-1} & \boldsymbol{X}_{k_{1}} & \cdots & \boldsymbol{X}_{n_{1}-1}
\end{array}\right],\label{eq:XeEnhancedForm}
\end{equation}
where each block is a $k_{2}\times\left(n_{2}-k_{2}+1\right)$ Hankel
matrix defined such that for all $0\leq l<n_{1}$:
\[
\boldsymbol{X}_{l}:=\left[\begin{array}{cccc}
x_{l,0} & x_{l,1} & \cdots & x_{l,n_{2}-k_{2}}\\
x_{l,1} & x_{l,2} & \cdots & x_{l,n_{2}-k_{2}+1}\\
\vdots & \vdots & \vdots & \vdots\\
x_{l,k_{2}-1} & x_{l,k_{2}} & \cdots & x_{l,n_{2}-1}
\end{array}\right].
\]
This form allows us to derive, through algebraic manipulation or a tensor product approach, the Vandemonde decomposition of each row of $\boldsymbol{X}$ as follows\begin{equation}
\forall l\text{ }(0\leq l<n_{1}):\quad\boldsymbol{X}_{l}=\boldsymbol{Z}_{\text{L}}\boldsymbol{Y}_{\text{d}}^{l}\boldsymbol{D}\boldsymbol{Z}{}_{\text{R}},\label{eq:XEachBlock_YZ}
\end{equation}
where $\boldsymbol{Z}_{\text{L}}$, $\boldsymbol{Z}_{\text{R}}$ and
$\boldsymbol{Y}_{\text{d}}$ are defined as
\[
\boldsymbol{Z}_{\text{L}}:=\left[\begin{array}{cccc}
1 & 1 & \cdots & 1\\
z_{1} & z_{2} & \cdots & z_{r}\\
\vdots & \vdots & \vdots & \vdots\\
z_{1}^{k_{2}-1} & z_{2}^{k_{2}-1} & \cdots & z_{r}^{k_{2}-1}
\end{array}\right],
\]
\[
\boldsymbol{Z}_{\text{R}}:=\left[\begin{array}{cccc}
1 & z_{1} & \cdots & z_{1}^{n_{2}-k_{2}}\\
1 & z_{2} & \cdots & z_{2}^{n_{2}-k_{2}}\\
\vdots & \vdots & \vdots & \vdots\\
1 & z_{r} & \cdots & z_{r}^{n_{2}-k_{2}}
\end{array}\right],
\]
and $\boldsymbol{Y}_{\text{d}}:=\text{diag}\left[y_{1},y_{2},\cdots,y_{r}\right]$.
Plugging (\ref{eq:XEachBlock_YZ}) into (\ref{eq:XeEnhancedForm})
yields the following:
\begin{equation*}
\boldsymbol{X}_{\text{e}}=\underset{\sqrt{k_1k_2}\boldsymbol{E}_{\text{L}}}{\underbrace{\left[\begin{array}{c}
\boldsymbol{Z}_{\text{L}}\\
\boldsymbol{Z}_{\text{L}}\boldsymbol{Y}_{\text{d}}\\
\vdots\\
\boldsymbol{Z}_{\text{L}}\boldsymbol{Y}_{\text{d}}^{k_{1}-1}\end{array}\right]}}\boldsymbol{D}\underset{\sqrt{(n_1-k_1+1)(n_2-k_2+1)}\boldsymbol{E}_{\text{R}}}{\underbrace{\left[\boldsymbol{Z}_{\text{R}},\boldsymbol{Y}_{\text{d}}\boldsymbol{Z}_{\text{R}},\cdots,\boldsymbol{Y}_{\text{d}}^{n_{1}-k_{1}}\boldsymbol{Z}_{\text{R}}\right]}},\end{equation*}
where $\boldsymbol{E}_\text{L}$ and $\boldsymbol{E}_\text{R}$  characterize the column and row space of $\boldsymbol{X}_{\text{e}}$,
respectively. The effectiveness of the enhanced form relies on the shift-invariance of harmonic structures (any $k$ consecutive samples lies in the same subspace), which promotes the use of Hankel matrices.

One can now see that $\boldsymbol{X_{\text{e}}}$ is low-rank or, $
\text{rank}\left(\boldsymbol{X}_{\text{e}}\right)\leq r$. We then attempt recovery via the following Enhanced Matrix Completion (EMaC) algorithm:
\begin{align}
\text{(EMaC)}:\quad\underset{\boldsymbol{M}\in\mathbb{C}^{n_{1}\times n_{2}}}{\text{minimize}}\quad & \left\Vert \boldsymbol{M}_{\text{e}}\right\Vert _{*}\\
\text{subject to}\quad &\mathcal{\mathcal{P}}_{\Omega}\left(\boldsymbol{M}\right)=\mathcal{\mathcal{P}}_{\Omega}\left(\boldsymbol{X}\right) \nonumber,
\end{align}
which minimizes the nuclear norm of the enhanced form over the constraint set. This convex program can be solved using off-the-shelf semidefinite program solvers in a tractable manner.

\subsection{Higher-Dimensional Frequency Model\label{sub:Extension-to-Higher-Dimension}}

The EMaC method extends to higher dimensional frequency models without
difficulty. For $K$-dimensional frequency models, one can convert
the original data to a $K$-fold Hankel matrix of rank at most $r$.
For instance, consider a 3 dimensional (3-D) model such that $x_{l_{1},l_{2},l_{3}}=\sum_{i=1}^{r}d_{i}y_{i}^{l_{1}}z_{i}^{l_{2}}w_{i}^{l_{3}}$.
An enhanced form can be defined as a 3-fold Hankel matrix such that
\[
\boldsymbol{X}_{\text{e}}:=\left[\begin{array}{cccc}
\boldsymbol{X}_{0,\text{e}} & \boldsymbol{X}_{1,\text{e}} & \cdots & \boldsymbol{X}_{n_{3}-k_{3},\text{e}}\\
\boldsymbol{X}_{1,\text{e}} & \boldsymbol{X}_{2,\text{e}} & \cdots & \boldsymbol{X}_{n_{3}-k_{3}+1,\text{e}}\\
\vdots & \vdots & \vdots & \vdots\\
\boldsymbol{X}_{k_{3}-1,\text{e}} & \boldsymbol{X}_{k_{1},\text{e}} & \cdots & \boldsymbol{X}_{n_{3}-1,\text{e}}
\end{array}\right],
\]
where $\boldsymbol{X}_{i,\text{e}}$ denotes the 2-D enhanced form
of the matrix consisting of all entries $x_{l_{1},l_{2},l_{3}}$ obeying
$l_{3}=i$. One can verify that $\boldsymbol{X}_{\text{e}}$ is of
rank at most $r$, and can thus apply EMaC on the 3-D enhanced
form $\boldsymbol{X}_{\text{e}}$. To summarize, for $K$-dimensional
frequency models, EMaC minimizes the nuclear norm over all
$K$-fold Hankel matrices consistent with the observed entries.

\subsection{Noisy Data}

In practice, measurements are always corrupted by a certain
amount of noise. To make our model and algorithm more practically
applicable, we can replace our measurements by $\boldsymbol{X}_{il}^{\text{o}}$
through the following noisy model
\[
\forall(i,l)\in\Omega:\quad\boldsymbol{X}_{il}^{\text{o}}=\boldsymbol{X}_{il}+\boldsymbol{N}_{il},
\]
where $\boldsymbol{X}_{il}^{\text{o}}$ is the observed $(i,l)$-th
entry, and $\boldsymbol{N}_{il}$ denotes the noise. Suppose that the noise satisfies $\left\Vert \mathcal{P}_{\Omega}\left(\boldsymbol{N}\right)\right\Vert _{\text{F}}\leq \delta$, 
then EMaC can be modified as: 
\begin{align}
\text{(EMaC-Noisy)}:\;\underset{\boldsymbol{M}\in\mathbb{C}^{n_{1}\times n_{2}}}{\text{minimize}}\quad & \left\Vert \boldsymbol{M}_{\text{e}}\right\Vert _{*}\label{eq:EMaCNoisy}\\
\text{subject to}\quad & \left\Vert \mathcal{\mathcal{P}}_{\Omega}\left(\boldsymbol{M}-\boldsymbol{X}^{\text{o}}\right)\right\Vert _{\text{F}}\leq\delta.\nonumber 
\end{align}

\section{Main Results \label{sec:Main-Results}}

Encouragingly, under certain incoherence conditions, the simple EMaC enables accurate recovery of the true data matrix from a small number of noiseless time-domain samples, and is stable against bounded noise. 

For convenience of presentation, we denote $\Omega_{\text{e}}(i,l)$ as the set of locations
in the enhanced matrix $\boldsymbol{X}_{\text{e}}$ containing copies
of $x_{i,l}$, and let $\omega_{i,l}:=\left|\Omega_{\text{e}}\left(i,l\right)\right|$. For each $\left(i,l\right)$, we use $\boldsymbol{A}_{(i,l)}$
to denote a {\em basis} matrix that extracts the average of all entries
in $\Omega_{\text{e}}\left(i,l\right)$, i.e.
\begin{equation*}
\left(\boldsymbol{A}_{(i,l)}\right)_{\alpha,\beta}:=\begin{cases}
\frac{1}{\sqrt{\left|\Omega_{\text{e}}\left(i,l\right)\right|}},&\text{if~}\left(\alpha,\beta\right)\in\Omega_{\text{e}}(i,l),\\
0, & \text{else}.
\end{cases}\end{equation*}

\subsection{Incoherence Measures}

In general, matrix completion from a few entries is hopeless unless
the underlying structure is uncorrelated with the observation basis.
This inspires us to define certain incoherence measures. Let $\boldsymbol{G}_{\text{L}}$
and $\boldsymbol{G}_{\text{R}}$ be $r\times r$ correlation matrices
such that for any $i\neq j$,
\small
\[
\left(\boldsymbol{G}_{\text{L}}\right)_{ij}:=\frac{1}{k_{1}k_{2}}\frac{1-\left(y_{i}^{*}y_{j}\right)^{k_{1}}}{1-y_{i}^{*}y_{j}}\frac{1-\left(z_{i}^{*}z_{j}\right)^{k_{2}}}{1-z_{i}^{*}z_{j}},
\]
\[
\left(\boldsymbol{G}_{\text{R}}\right)_{ij}:=\frac{1-\left(y_{i}^{*}y_{j}\right)^{n_{1}-k_{1}+1}}{\left(n_{1}-k_{1}+1\right)(1-y_{i}^{*}y_{j})}\frac{1-\left(z_{i}^{*}z_{j}\right)^{n_{2}-k_{2}+1}}{\left(n_{2}-k_{2}+1\right)(1-z_{i}^{*}z_{j})},
\]
\normalsize
with the convention $\left(\boldsymbol{G}_{\text{L}}\right)_{ii}=\left(\boldsymbol{G}_{\text{R}}\right)_{ii}=1$. Note that $\boldsymbol{G}_{\text{L}}$ and $\boldsymbol{G}_{\text{R}}$ can be obtained by sampling the 2-D Dirichlet kernel, which is frequently used in Fourier analysis. Our incoherence measure is defined as follows.

\begin{definition}[\textbf{Incoherence}]Let $\boldsymbol{X}_{\mathrm{e}}$
denote the enhanced matrix associated with $\boldsymbol{X}$, and
suppose the SVD of $\boldsymbol{X}_{\mathrm{e}}$ is given by $\boldsymbol{X}_{\mathrm{e}}=\boldsymbol{U}{\bm\Lambda}\boldsymbol{V}^{*}$.
Then $\boldsymbol{X}$ is said to have incoherence $\left(\mu_{1},\mu_{2},\mu_{3}\right)$
if $ $they are respectively the smallest quantities such that 
\begin{equation}
\sigma_{\min}\left(\boldsymbol{G}_{\mathrm{L}}\right)\geq 1/\mu_{1}, \quad\sigma_{\min}\left(\boldsymbol{G}_{\mathrm{R}}\right)\geq 1/\mu_{1};\label{eq:LeastSV_G}
\end{equation}
\begin{equation}
\max_{(i,l)\in[n_{1}]\times[n_{2}]}\frac{\Big|\sum_{(\alpha,\beta)\in\Omega_{\mathrm{e}}\left(i,l\right)}\left(\boldsymbol{U}\boldsymbol{V}^{*}\right)_{\alpha,\beta}\Big|^{2}}{\left|\Omega_{\mathrm{e}}(i,l)\right|^{2}}\leq\frac{\mu_{2}r}{n_{1}^{2}n_{2}^{2}};\label{eq:Inhomogenuity_UV}
\end{equation}
and
\begin{equation}
\sum_{\boldsymbol{a}\in[n_{1}]\times[n_{2}]}\left|\left\langle \boldsymbol{U}\boldsymbol{U}^{*}\boldsymbol{A}_{b}\boldsymbol{V}\boldsymbol{V}^{*},\sqrt{\omega_{a}}\boldsymbol{A}_{\boldsymbol{a}}\right\rangle \right|^{2}\leq\frac{\mu_{3}r}{n_{1}n_{2}}\omega_{\boldsymbol{b}}\label{eq:APtAHomogenuity}
\end{equation}
holds for all $\boldsymbol{b}\in[n_{1}]\times[n_{2}]$. \end{definition}

Some brief interpretations of the above incoherence conditions are
in order:

Condition \eqref{eq:LeastSV_G} specifies certain incoherence among the locations of
frequency pairs, which does not coincide with and is not subsumed by the separation condition
required in \cite{CandesFernandez2012SR,TangBhaskarShahRecht2012}.
The frequency pairs can be spread out (e.g. when their locations are
generated in some random fashion), or minimally separated (e.g. when
they are small perturbation of a fine grid). 

Condition \eqref{eq:Inhomogenuity_UV} can be satisfied when the total energy of
each skew diagonal of $\boldsymbol{U}\boldsymbol{V}^{*}$ is proportional
to the dimension of this skew diagonal. This is indeed a weaker condition than the one introduced in \cite{ExactMC09} for matrix completion, which requires uniform energy distribution over all entries of $\boldsymbol{U}\boldsymbol{V}^{*}$. For instance, an ideal $\mu_2$ can often be obtained when the complex phase of all frequencies are generated in some random fashion. 

Condition \eqref{eq:APtAHomogenuity} is an incoherence measure based on the ($K$-fold) Hankel structures. For example, one can reason that a desired $\mu_3$ can be obtained if the magnitude of all entries
 of $\boldsymbol{U}\boldsymbol{U}^{*}\boldsymbol{A}_{a}\boldsymbol{V}\boldsymbol{V}^{*}$
is mostly even. Condition \eqref{eq:LeastSV_G} and \eqref{eq:APtAHomogenuity} depend
only on the frequency pairs. In fact, one can verify that $\boldsymbol{U}\boldsymbol{U}^{*}=\boldsymbol{E}_{\text{L}}\left(\boldsymbol{E}_{\text{L}}^{*}\boldsymbol{E}_{\text{L}}\right)^{-1}\boldsymbol{E}_{\text{L}}^{*}$
and $\boldsymbol{V}\boldsymbol{V}^{*}=\boldsymbol{E}_{\text{R}}^{*}\left(\boldsymbol{E}_{\text{R}}\boldsymbol{E}_{\text{R}}^{*}\right)^{-1}\boldsymbol{E}_{\text{R}}$,
which depend only on the locations of the frequency pairs. Condition
(\ref{eq:Inhomogenuity_UV}), however, might also rely on the amplitudes $\boldsymbol{D}$. 

Finally, the incoherence measures $\left(\mu_{1},\mu_{2},\mu_{3}\right)$
are mutually correlated, which is supplied as follows.
\begin{lemma}\label{lemma-UpperBoundAbPtAa}The incoherence measures $(\mu_{1},\mu_{2},\mu_{3})$ of $\boldsymbol{X}_{\text{e}}$ satisfy
\begin{equation}
\mu_{2}\leq\mu_{1}^{2}c_{\text{s}}^{2}r,\quad\text{and}\quad\mu_{3}\leq\mu_{1}^{2}c_{\text{s}}^{2}r.\label{eq:Mu2BoundViaMu1}
\end{equation}
\end{lemma} 

\subsection{Theoretical Guarantees}
With the above incoherence measures, the main theoretical guarantee
is supplied in the following theorem. \begin{theorem}\label{theorem-EMaC-noiseless}Let
$\boldsymbol{X}$ be a data matrix with matrix form (\ref{eq:X_MatrixForm}),
and $\Omega$ the random location set of size $m$. Define $c_{\mathrm{s}}:=\max\left(\frac{n_{1}n_{2}}{k_{1}k_{2}},\frac{n_{1}n_{2}}{\left(n_{1}-k_{1}+1\right)\left(n_{2}-k_{2}+1\right)}\right)$.
If all measurements are noiseless, then there exists a constant $c_{1}>0$
such that under either of the following conditions:
\begin{itemize}
\item Conditions \eqref{eq:LeastSV_G}, \eqref{eq:Inhomogenuity_UV} and \eqref{eq:APtAHomogenuity} hold and
\begin{equation} \label{firsthalf}
m>c_{1}\max\left(\mu_{1}c_{\mathrm{s}},\mu_{2},\mu_{3}c_{\mathrm{s}}\right)r\log^{2}\left(n_{1}n_{2}\right);
\end{equation}
\item Condition \eqref{eq:LeastSV_G} holds and 
\begin{equation} \label{secondhalf}
m>c_{1}\mu_{1}^{2}c_{\mathrm{s}}^{2}r^{2}\log^{2}(n_{1}n_{2});
\end{equation}
\end{itemize}
$\boldsymbol{X}$ is the unique solution of EMaC with probability
exceeding $1-(n_{1}n_{2})^{-2}$. \end{theorem}

Note that \eqref{secondhalf}
is an immediate consequence of \eqref{firsthalf} by Lemma~\ref{lemma-UpperBoundAbPtAa}. Theorem \ref{theorem-EMaC-noiseless} states the following: (1) under
\emph{strong} incoherence condition (i.e. given that $\left(\mu_{1},\mu_{2},\mu_{3}\right)$
are all constants), prefect recovery is possible as soon as the number
of measurements exceeds the order of $r\log^{2}\left(n_{1}n_{2}\right)$;
(2) under \emph{weak} incoherence condition (i.e. given only that
$\mu_{1}$ is a constant), perfect recovery is possible from $\mathcal{O}\left(r^{2}\log\left(n_{1}n_{2}\right)\right)$
samples. Since there are at least $\Theta(r)$ degrees
of freedom in total, this establishes the near optimality of EMaC
under strong incoherence condition.

We would like to note that while we assume random observation models, the conditions imposed on the data model are deterministic. This is different from \cite{TangBhaskarShahRecht2012}, where randomness are assumed for both the observation model and the data model. 

On the other hand, our method enables stable recovery even when the time-domain samples are noisy copies of the true data. Here, we say the recovery is stable if the solution of EMaC-Noisy is ``close'' to the ground truth. To this end, we establish the following theorem, which is a counterpart of Theorem \ref{theorem-EMaC-noiseless} in the noisy setting. 

\begin{theorem} \label{theorem-EMaC-noisy} Suppose $\boldsymbol{X}^{\mathrm{o}}$ is a noisy copy of $\boldsymbol{X}$ that satisfies $\|\mathcal{P}_\Omega(\boldsymbol{X}-\boldsymbol{X}^{\mathrm{o}})\|_{\mathrm{F}}\leq \delta$.  Under the conditions of Theorem~\ref{theorem-EMaC-noiseless}, the solution to EMaC-Noisy in \eqref{eq:EMaCNoisy} satisfies
\begin{equation} \| \hat{\boldsymbol{X}}_{\mathrm{e}} - \boldsymbol{X}_{\mathrm{e}} \|_{\mathrm{F}} \leq \left\{2\sqrt{n_1n_2}+8n_1n_2+\frac{8\sqrt{2}n_1^2n_2^2}{m}\right\} \delta\nonumber\end{equation}
with probability exceeding $1-(n_1n_2)^{-2}$.
\end{theorem}
Theorem~\ref{theorem-EMaC-noisy} basically implies that the recovered enhanced matrix (which contains $\mathcal{O}(n_1^2n_2^2)$ entries) is close to the true enhanced matrix at high signal-to-noise ratio. In particular, the average entry inaccuracy is bounded above by $\mathcal{O}(\frac{n_1n_2}{m}\delta)$. We note that in practice, EMaC-Noisy usually yields better estimate, possibly by a polynomial factor. The practical applicability will be illustrated in  Section \ref{sec:Numerical-Experiments} through numerical examples. 

\section{Numerical Experiments \label{sec:Numerical-Experiments}}

\subsection{Phase Transition for Exact Recovery \label{sec:Numerical-Exact-Recovery}}
We examine phase transition of the EMaC algorithm to evaluate its practical ability. A square enhanced form was adopted with $n_{1}=n_{2}$, which corresponds to the smallest $c_\text{s}$. For each $(r,m)$ pair, we generated a spectrally
sparse data matrix $\boldsymbol{X}$ by randomly generating $r$
frequency spikes in $[0,1]\times[0,1]$, and sampled a subset $\Omega$ of size $m$
entries uniformly at random. The EMaC algorithm was conducted
using CVX with SDPT3. Each trial is declared successful if the normalized
mean squared error $\|\hat{\boldsymbol{X}}-\boldsymbol{X}\|_{\text{F}}/\|\boldsymbol{X}\|_{\text{F}}\leq10^{-4}$, where $\hat{\boldsymbol{X}}$ denotes the estimate obtained through
EMaC. The empirical success rate is calculated by averaging over 100 Monte Carlo trials.

\begin{figure}
\centering\includegraphics[width=0.45\textwidth]{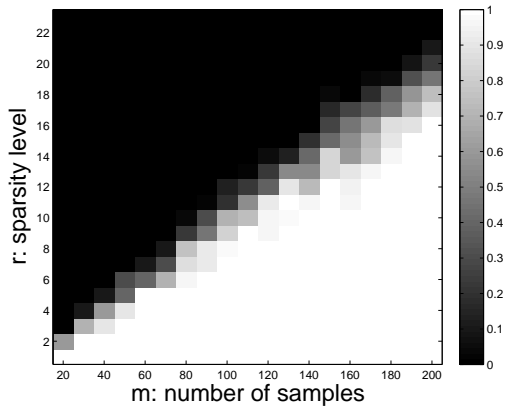}
\caption{\label{fig:Phase-transition-plots-Hankel2D}Phase transition diagrams
where spike locations are randomly generated. The results are shown for the case where $n_{1}=n_{2}=15$.}
\end{figure}

Fig. \ref{fig:Phase-transition-plots-Hankel2D} illustrates the results
of these Monte Carlo experiments when $\boldsymbol{X}$ is a $15\times15$ matrix. The empirical success rate is reflected by the color of each cell. It can be seen
that the number of samples $m$ grows approximately linearly with respect to the spectral sparsity $r$, in line with our theoretical guarantee in Theorem~\ref{theorem-EMaC-noiseless}. This phase transition diagram validates the practical
applicability of our algorithm in the noiseless setting.

%

\subsection{Stable Recovery via Singular Value Thresholding}
The above experiments were conducted using the advanced
semidefinite programming solver SDPT3. This and many other popular
solvers (like SeDuMi) are based on interior point methods, which are
typically inapplicable to large-scale data. In fact, SDPT3 fails to
handle an $n\times n$ data matrix when $n$ exceeds $19$, which
corresponds to a $100\times100$ enhanced matrix. 

One solution for large-scale data is the first-order algorithms tailored
for MC problems, e.g. the singular value thresholding
(SVT) algorithm developed in \cite{cai2010singular}. We propose a modified SVT algorithm in Algorithm \ref{alg:SVT} to exploit the Hankel structure. 


\begin{algorithm}[h]
   \caption{Singular Value Thresholding for EMaC}
   \label{alg:SVT}
\begin{algorithmic}
\STATE {\bfseries Input:} The observed data matrix $\boldsymbol{X}^{\text{o}}$ on the location set $\Omega$.
   \STATE {\bfseries initialize} let
$\boldsymbol{X}_{\text{e}}^{\text{o}}$ denote the enhanced form of
$\mathcal{P}_{\Omega}\left(\boldsymbol{X}^{\text{o}}\right)$; set
$\boldsymbol{M}_{0}=\boldsymbol{X}_{\text{e}}^{\text{o}}$ and $t=0$.
   \REPEAT
   \STATE 1) $\boldsymbol{Q}_{t}\leftarrow\mathcal{D}_{\tau_{t}}\left(\boldsymbol{M}_{t}\right)$
   \STATE 2) $\boldsymbol{M}_{t}\leftarrow \mathcal{H}_{\boldsymbol{X}^\text{o}}(\boldsymbol{Q}_t)$
   \STATE 3) $t\leftarrow t+1$
   \UNTIL{convergence}
   \STATE {\bf output} $\hat{\boldsymbol{X}}$ as the matrix with enhanced form $\boldsymbol{M}_{t}$.
\end{algorithmic}
\end{algorithm}

In particular, $\mathcal{D}_{\tau_{t}}(\cdot)$ in Algorithm ~\ref{alg:SVT} denotes singular value shrinkage
operator. Specifically, if the SVD of $\boldsymbol{X}$ is given by
$\boldsymbol{X}=\boldsymbol{U}\boldsymbol{\Sigma}\boldsymbol{V}^{*}$
with $\boldsymbol{\Sigma}=\text{diag}\left(\left\{ \sigma_{i}\right\} \right)$,
then
\[
\mathcal{D}_{\tau_{t}}\left(\boldsymbol{X}\right):=\boldsymbol{U}\text{diag}\left(\left\{ \left(\sigma_{i}-\tau_{t}\right)_{+}\right\} \right)\boldsymbol{V}^{*}.
\]
where $\tau_t>0$ is the soft-thresholding level\footnote{A good soft thresholding level is hard to pick, and needs to be selected by cross validation. Hence, the phase transisions of SVT and EMaC do not necessarily coincide.}. Besides, in the $K$-dimensional frequency model, $\mathcal{H}_{\boldsymbol{X}^\text{o}}(\boldsymbol{Q}_t)$ denotes the projection of $\boldsymbol{Q}_t$ onto the subspace of enhanced matrices (i.e. $K$-fold Hankel matrices) that are consistent with the observed entries. Consequently, at each iteration, a pair $\left(\boldsymbol{Q}_{t},\boldsymbol{M}_{t}\right)$ is produced by first performing singular value shrinkage and
then projecting the outcome onto the space of $K$-fold Hankel matrices that are consistent with observed
entries. 
\begin{figure}
\hspace{-0.1in}\includegraphics[width=0.52\textwidth]{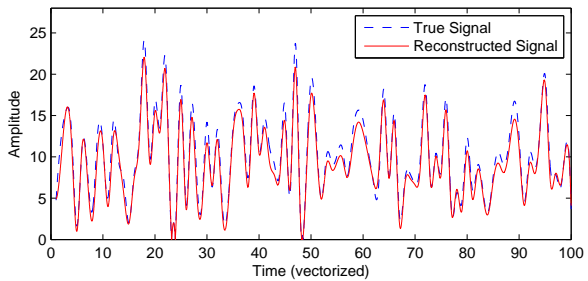}
\caption{\label{fig:SVTNoisy}The performance of SVT for a $101\times101$
data matrix that contains $30$ random frequency spikes. 
$5.8\%$ of all entries ($m=600$) are observed with signal-to-noise
amplitude ratio $10$. Here, $\tau_{t}=0.1\sigma_{\max}\left(\boldsymbol{M}_{t}\right)/\left\lceil \frac{t}{10}\right\rceil $
empirically. The reconstruction against the true
data for the first 100 time instances (after vectorization) are plotted.}
\end{figure}

Fig. \ref{fig:SVTNoisy} illustrates the performance of Algorithm~\ref{alg:SVT}. We generated a $101\times101$ data matrix $\boldsymbol{X}$ with a superposition of $30$ random complex
sinusoids, and revealed 5.8\% of the total entries (i.e. $m=600$)
uniformly at random. The noise was i.i.d. Gaussian-distributed
giving a signal-to-noise amplitude ratio of $10$. The reconstructed signal is superimposed on the ground truth in Fig. \ref{fig:SVTNoisy}. The normalized reconstruction
error was $\Vert \hat{\boldsymbol{X}}-\boldsymbol{X}\Vert _{\text{F}}/\left\Vert \boldsymbol{X}\right\Vert _{\text{F}}=0.1098$, validating the stability of EMaC in the presence of noise.

\subsection{ Super Resolution}
\begin{figure*}[htp]
\begin{tabular}{ccc}
\hspace{-0.3in} \includegraphics[height=2in, width=0.35\textwidth]{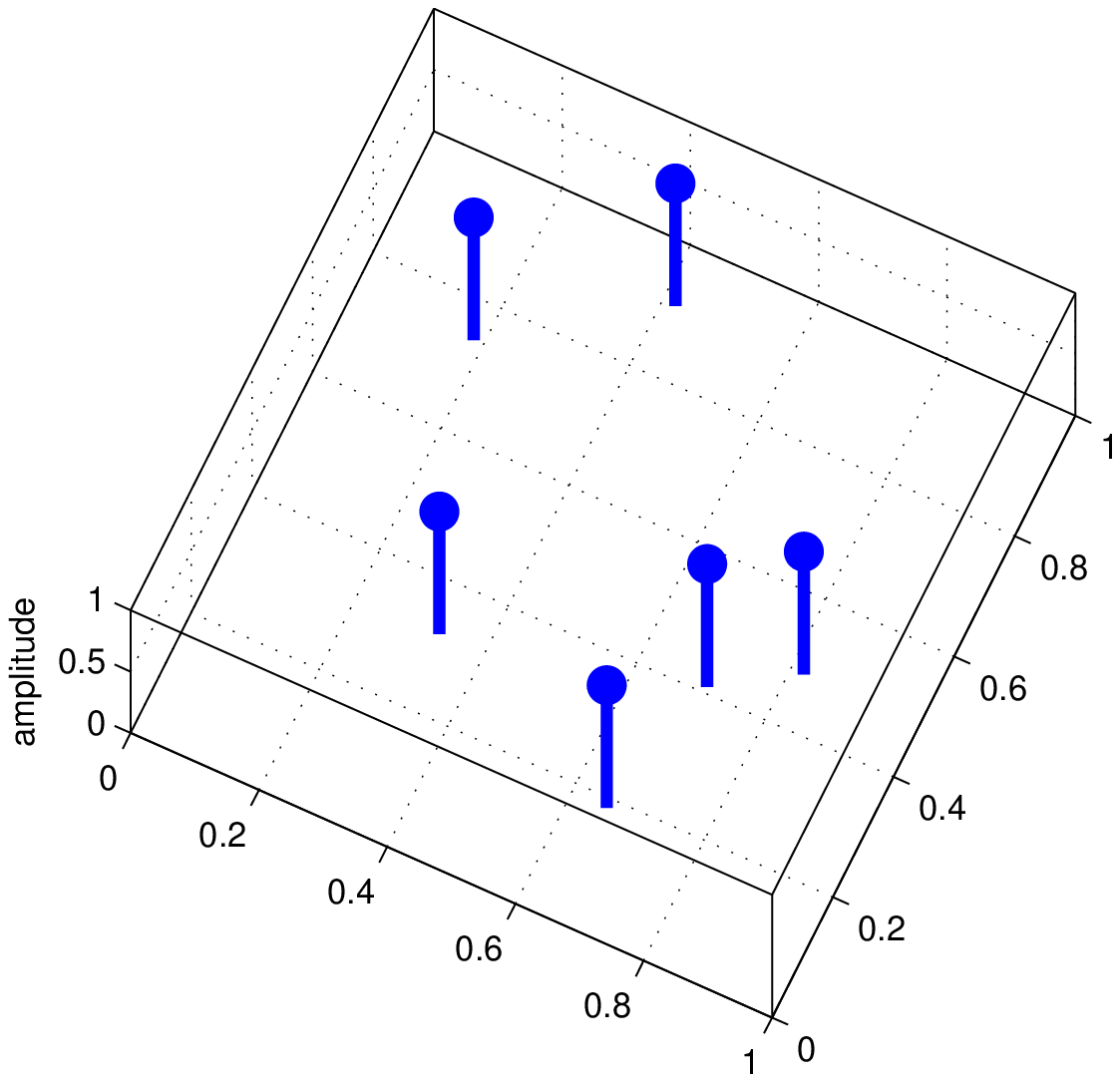}  & \hspace{-0.25in} \includegraphics[height=1.8in, width=0.35\textwidth]{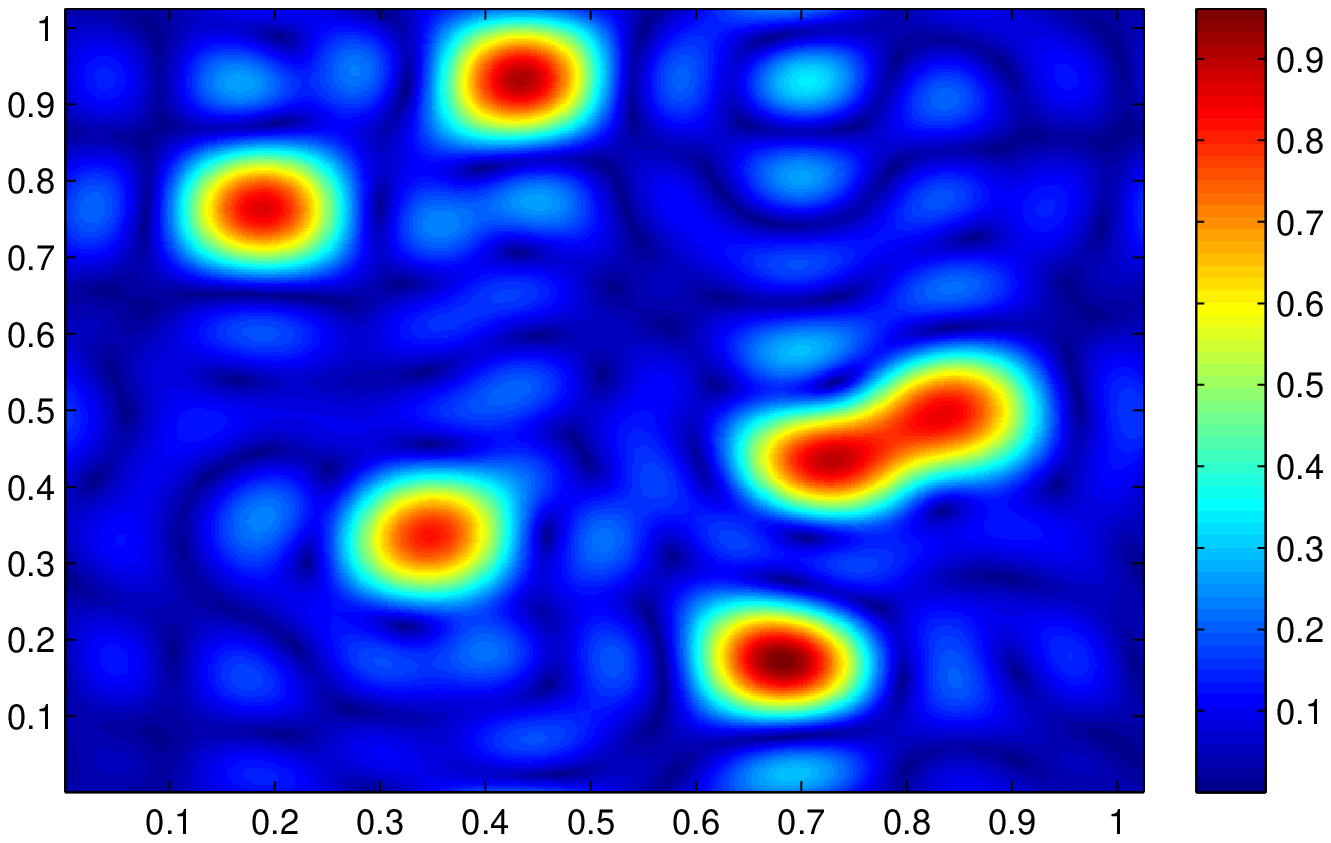}  & \hspace{-0.2in} \includegraphics[height=1.8in, width=0.32\textwidth]{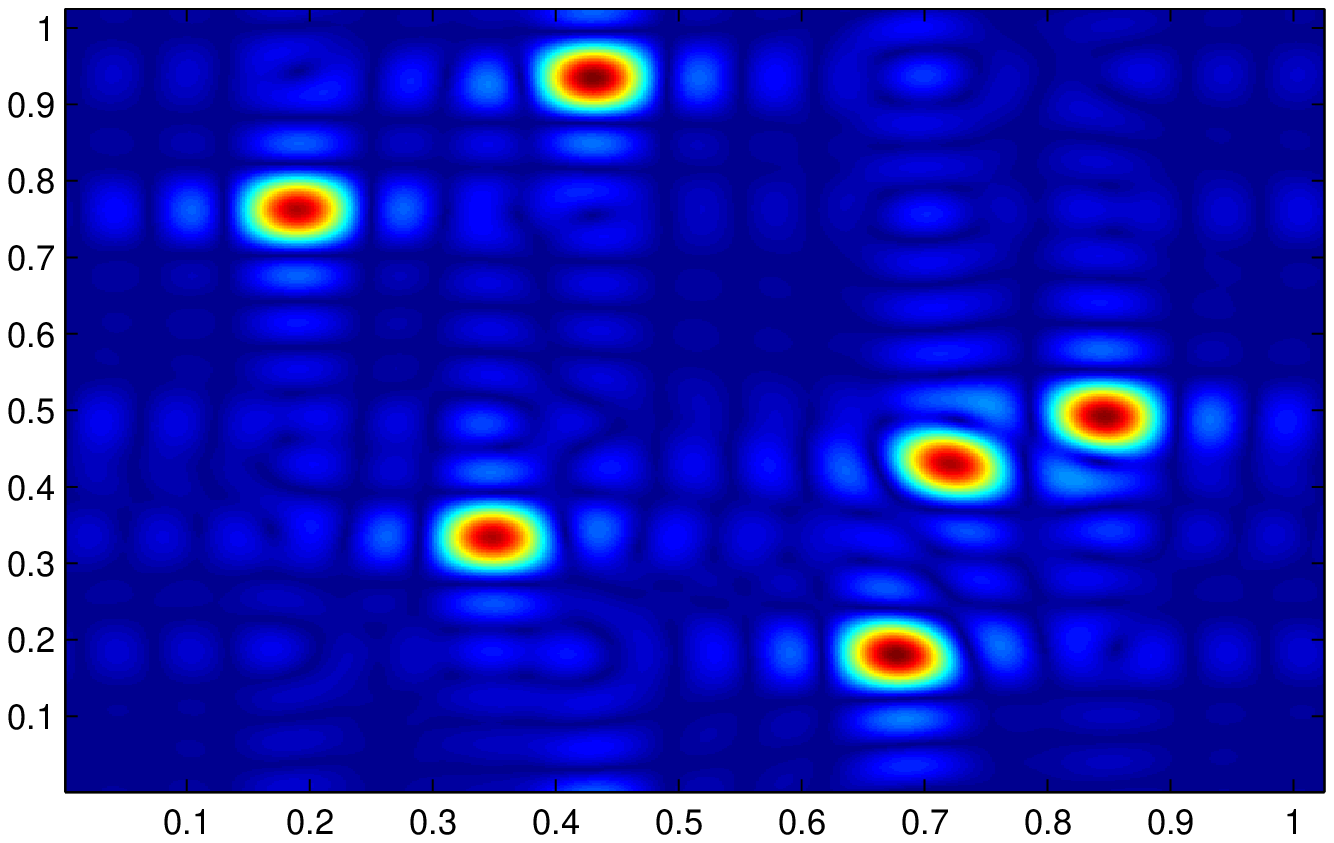} \tabularnewline
\hspace{-0.4in}(a) Ground truth  & (b) Low-resolution observation  & (c) High-resolution reconstruction \tabularnewline
\end{tabular}\caption{\label{fig:super_resolution}A synthetic super resolution example,
where the observation (b) is taken from low-frequency components of the
ground truth in (a), and the reconstruction (c) is done via inverse Fourier transform of extrapolated high-frequency components.}
\end{figure*}

The proposed EMaC algorithm works beyond the random observation model in Theorem~\ref{theorem-EMaC-noiseless}. Fig.~\ref{fig:super_resolution} considers a synthetic super resolution example, where the ground truth in Fig.~\ref{fig:super_resolution} (a) contains $6$ point sources with constant amplitude. The low-resolution observation in Fig.~\ref{fig:super_resolution} (b) is obtained by measuring low-frequency components $[-f_{\mathrm{lo}},f_{\mathrm{lo}}]$ of the ground truth. Due to the large width of the associated point-spread function, both the locations and amplitudes of the point sources are distorted in the low-resolution image. 

We apply EMaC to extrapolate high-frequency components up to $[-f_{\mathrm{hi}},f_{\mathrm{hi}}]$, where $f_{\mathrm{hi}}/f_{\mathrm{lo}}=2$. The reconstruction in Fig.~\ref{fig:super_resolution} (c) is obtained via applying directly inverse Fourier transform of the spectrum to avoid parameter estimation such as the number of modes. The resolution is greatly enhanced from Fig.~\ref{fig:super_resolution} (b), suggesting that EMaC is a promising approach for super resolution tasks.

\section{Structured Matrix Completion}\label{sub:Extension-to-Hankel}

One problem closely related to our method is completion of $K$-fold Hankel matrices from a small number of entries. Consider for
instance the 2-D model. While each spectrally sparse signal can be
mapped to a low-rank $2$-fold Hankel matrix, it is not clear whether all $2$-fold Hankel matrices
of rank $r$ can be written as the enhanced form of an object with
spectral sparsity $r$. Therefore, one can think of recovery of $K$-fold Hankel matrices as a more general problem than the spectral compressed sensing problem. Indeed, Hankel matrix completion has found numerous applications
in system identification \cite{Fazel2011hankel},
natural language processing \cite{Balle2012spectral}, computer vision \cite{Sankaranarayanan2010compressive}, medical imaging \cite{LustigHankel2013}, etc. 

There has been several work concerning algorithms and numerical experiments for Hankel matrix
completions \cite{Fazel2003Hankel,Fazel2011hankel,Markovsky2008structured}.
However, to the best of our knowledge, there has been little theoretical
guarantee that addresses directly Hankel matrix completion. Our analysis framework in Theorem \ref{theorem-EMaC-noiseless} can be straightforwardly adapted to the general $K$-fold Hankel matrix completions. Notice that $\mu_2$ and $\mu_3$ are defined using the SVD of $\boldsymbol{X}_{\text{e}}$ in \eqref{eq:Inhomogenuity_UV} and \eqref{eq:APtAHomogenuity}, and we only need to modify the definition of $\mu_1$, as stated in the following theorem.

\begin{theorem}\label{theorem-EMaC-Hankel}Consider a $K$-fold Hankel
matrix $\boldsymbol{X}_{\text{e}}$ of rank $r$. The bounds in Theorem
\ref{theorem-EMaC-noiseless} and Theorem
\ref{theorem-EMaC-noisy}
continue to hold, if the incoherence $\mu_{1}$ is defined as the
smallest quantity that satisfies: $\forall\left(i,l\right)\in[n_{1}]\times[n_{2}],$
\begin{equation}\label{eq:IncohrenceUU_Hankel}
\max\left\{\Vert \boldsymbol{U}\boldsymbol{U}^{*}\boldsymbol{A}_{(i,l)}\Vert _{\text{F}}^{2},\Vert \boldsymbol{A}_{(i,l)}\boldsymbol{V}\boldsymbol{V}^{*}\Vert _{\text{F}}^{2} \right\} \leq\frac{\mu_{1}c_{\text{s}}r}{n_{1}n_{2}}.
\end{equation}
\end{theorem}


Condition \eqref{eq:IncohrenceUU_Hankel} requires that the left and
right singular vectors are sufficiently uncorrelated with the observation
basis. In fact, condition \eqref{eq:IncohrenceUU_Hankel} is a much
weaker assumption than \eqref{eq:LeastSV_G}. 

It is worth mentioning that low-rank Hankel matrices can often
be converted to low-rank Toeplitz counterparts.
Both Hankel and Toeplitz matrices are important forms that capture
the underlying harmonic structures. Our results and analysis framework easily extend to the Toeplitz matrix completion problem.

\section{Conclusions\label{sec:conclusions-and-future}}
We present an efficient nonparametric algorithm to estimate a spectrally sparse object from its partial time-domain samples, which poses spectral compressed sensing as a low-rank Hankel structured matrix completion problem. Under mild conditions, our algorithm enables recovery of the multi-dimensional unknown frequencies with infinite precision, which remedies the basis mismatch issue that arises in conventional CS paradigms. To the best of our knowledge, our result on Hankel matrix completion is also the first theoretical guarantee that is close to the information-theoretical limit (up to a logarithmic factor). 



\section*{Appendix: Proof Outline of Theorem~\ref{theorem-EMaC-noiseless}\label{sec:Main-Proof}}

The EMaC method has similar spirit as the well-known matrix
completion algorithms \cite{ExactMC09,Gross2011recovering}, however
the additional Hankel or block-Hankel structures on the matrices make existing theoretical results inapplicable in our framework. Nevertheless, the golfing scheme introduced in \cite{Gross2011recovering} lays the foundation of our analysis. We provide here a sketch of the proof, with detailed derivation deferred to \cite{ChenChi2013TIT} and the supplemental material.


Denote by $T$ the tangent space with respect to $\boldsymbol{U}$ and
$\boldsymbol{V}$. Let $\mathcal{P}_{T}$ be the orthogonal projection
onto $T$, with $\mathcal{P}_{T^\perp}:=\mathcal{I}-\mathcal{P}_T$ denoting its orthogonal complement.
Denote by $\mathcal{A}_{\left(i,l\right)}$ the orthogonal projection
onto the subspace spanned by $\boldsymbol{A}_{(i,l)}$. The projection
onto the space spanned by all $\boldsymbol{A}_{(i,l)}$'s and
its complement are defined as 
\begin{equation}
\mathcal{A}:=\sum_{\left(i,l\right)\in[n_{1}]\times[n_{2}]}\mathcal{A}_{(i,l)},\quad\text{and}\quad\mathcal{A}^{\perp}=\mathcal{I}-\mathcal{A}.
\end{equation}
Suppose that $\Omega$ is obtained by sampling with replacement, i.e. 
$\Omega$ contains $m$ indices $\{z_i\}_{i=1}^{m}$ that are i.i.d. generated from $[n_{1}]\times[n_{2}]$. We define the associated projection operators
as $\mathcal{A}_{\Omega}:=\sum_{i=1}^{m}\mathcal{A}_{z_{i}}$, and
introduce another projection operator $\mathcal{A}^{\prime}_{\Omega}$ similar to $\mathcal{A}_{\Omega}$ but with the summation only over distinct samples.

To prove exact recovery of EMaC, it is sufficient to produce
a dual certificate as follows.
\begin{lemma}\label{lemma-Dual-Certificate}For a location set $\Omega$
of size $m$, suppose that the sampling operator
$\mathcal{P}_{\Omega}$ obeys 
\begin{equation}
\left\Vert \mathcal{P}_{T}\mathcal{A}\mathcal{P}_{T}-\frac{n_{1}n_{2}}{m}\mathcal{P}_{T}\mathcal{A}_{\Omega}\mathcal{P}_{T}\right\Vert \leq\frac{1}{2}.\label{eq:WellConditionPtAomegaPt}
\end{equation}
If there exists a matrix $\boldsymbol{W}$ that obeys
\begin{equation}
\begin{cases}
\left(\mathcal{A}-\mathcal{A}^{\prime}_{\Omega}\right)\left(\boldsymbol{U}\boldsymbol{V}^{*}+\boldsymbol{W}\right)=0,\\
\left\Vert \mathcal{P}_{T}\left(\boldsymbol{W}\right)\right\Vert _{\mathrm{F}}\leq\frac{1}{2n_{1}^{2}n_{2}^{2}},\\
\left\Vert \mathcal{P}_{T^{\perp}}\left(\boldsymbol{W}\right)\right\Vert \leq\frac{1}{2},\label{eq:UV_{W}{}_{C}ontained_{i}n_{A}Omega}
\end{cases}
\end{equation}
then $\boldsymbol{X}_{\text{e}}$ is the unique optimizer of EMaC.\end{lemma}




The dual certificate is constructed via the golfing scheme introduced
in \cite{Gross2011recovering}. Specifically, we generate $j_{0}$
independent random location sets $\Omega_{j}\sim\text{Bernoulli}(q)$, which is sampled with replacement. Set $q$ such that $1-\rho=\left(1-q\right)^{j_{0}}$, where $\rho:=\frac{m}{n_{1}n_{2}}$, then the distribution of $\Omega$ is the same as $\Omega_{1}\cup\Omega_{2}\cup\cdots\cup\Omega_{j_{0}}$. For small
$\rho$, we have $q=\mathcal{O}\left(\rho/j_{0}\right)$ through the
Taylor expansion. Consider a small constant $\epsilon<\frac{1}{3}$. The construction of a dual certificate $\boldsymbol{W}$ proceeds as follows:
\begin{enumerate}
\item Set $\boldsymbol{B}_{0}=0$,
and $j_{0}:=3\log_{1/\epsilon}(n_{1}n_{2})$.
\item For $1\leq i\leq j_{0}$, let 
$$\boldsymbol{B}_{i}=\boldsymbol{B}_{i-1}+\left(q^{-1}\mathcal{A}_{\Omega_{i}}+\mathcal{A}^{\perp}\right)\mathcal{P}_{T}\left(\boldsymbol{U}\boldsymbol{V}^{*}-\boldsymbol{B}_{i-1}\right).$$ 
\item Set $\boldsymbol{W}:=-\left(\boldsymbol{U}\boldsymbol{V}^{*}-\boldsymbol{B}_{j_{0}}\right)$.
\end{enumerate}

We will verify that $\boldsymbol{W}$ is a valid dual certificate satisfying the
conditions \eqref{eq:UV_{W}{}_{C}ontained_{i}n_{A}Omega} in Lemma \ref{lemma-Dual-Certificate}. Our construction immediately yields $\left(\mathcal{A}'_{\Omega}+\mathcal{A}^{\perp}\right)\left(\boldsymbol{B}_{i}\right)=\boldsymbol{B}_{i}$ for $1\leq i\leq j_{0}$, which gives
\[
\left(\mathcal{A}-\mathcal{A}{}_{\Omega}^\prime\right)\left(\boldsymbol{U}\boldsymbol{V}^{*}+\boldsymbol{W}\right)=0.
\]
Define $\boldsymbol{F}_{i}:=\boldsymbol{U}\boldsymbol{V}^{*}-\boldsymbol{B}_{i},$
and hence $\boldsymbol{W}=\boldsymbol{F}_{j_{0}}$, one has\begin{align*}
\mathcal{P}_{T}\left(\boldsymbol{F}_{i}\right) & =\left(\mathcal{P}_{T}\mathcal{A}\mathcal{P}_{T}-q^{-1}\mathcal{P}_{T}\mathcal{A}_{\Omega_{i}}\mathcal{P}_{T}\right)\left(\boldsymbol{F}_{i-1}\right).
\end{align*}
To proceed, we present Lemma \ref{lemma-Invertibility-PtWPt} which shows that $\mathcal{A}_{\Omega}$
is sufficiently incoherent with respect to $T$. 
\begin{lemma}\label{lemma-Invertibility-PtWPt}There exists a constant
$c_{1}>0$ such that if $m>c_{1}\mu_{1}c_{\text{s}}r\log\left(n_{1}n_{2}\right)$,
then
\begin{equation}
\left\Vert \frac{n_{1}n_{2}}{m}\mathcal{P}_{T}\mathcal{A}_{\Omega}\mathcal{P}_{T}-\mathcal{P}_{T}\mathcal{A}\mathcal{P}_{T}\right\Vert \leq{\epsilon}\label{eq:Invertibility-PtWPt}
\end{equation}
with probability exceeding $1-\left(n_{1}n_{2}\right)^{-4}$.\end{lemma}
Under the assumptions of Lemma~\ref{lemma-Invertibility-PtWPt}, we have
\[
\left\Vert \mathcal{P}_{T}\left(\boldsymbol{W}\right)\right\Vert _{\text{F}}=\left\Vert \mathcal{P}_{T}\left(\boldsymbol{F}_{j_{0}}\right)\right\Vert _{\text{F}}\leq\epsilon^{j_{0}}\sqrt{r}<{n_{1}^{-2}n_{2}^{-2}}.
\]
It remains to show $\left\Vert \mathcal{P}_{T^{\perp}}\left(\boldsymbol{W}\right)\right\Vert \leq\frac{1}{2}$ as follows.
\begin{lemma}\label{lemma-Pt_perp-F}There exist constants $c_{8},c_{9}>0$ such that if $m>c_{9}\max\left(\mu_{1}c_{\text{s}},\mu_{2},\mu_{3}\right)r\log^{2}\left(n_{1}n_{2}\right)$, then
\[
\forall1\leq i\leq j_{0}:\;\left\Vert \mathcal{P}_{T^{\perp}}\left(q^{-1}\mathcal{A}_{\Omega_{i}}+\mathcal{A}^{\perp}\right)\mathcal{P}_{T}\left(\boldsymbol{F}_{i}\right)\right\Vert <2^{-i-2}
\]
with probability at least $1-c_{8}(n_{1}n_{2})^{-3}$.
\end{lemma}

Under the assumptions of Lemma \ref{lemma-Pt_perp-F}, we have \begin{align*}\left\Vert \mathcal{P}_{T^{\perp}}\left(\boldsymbol{W}\right)\right\Vert  & \leq\sum_{i=0}^{j_{0}}\left\Vert \mathcal{P}_{T^{\perp}}\left(q^{-1}\mathcal{A}_{\Omega}+\mathcal{A}^{\perp}\right) \mathcal{P}_{T}\left(\boldsymbol{F}_{i}\right)\right\Vert < 1/2.
\end{align*}
So far, we have successfully shown that $\boldsymbol{W}$ is a valid
dual certificate with high probability, and hence EMaC allows
exact recovery with high probability. 

\bibliography{bibfileSparseMatrixPencil}

\begin{thebibliography}{27}
\providecommand{\natexlab}[1]{#1}
\providecommand{\url}[1]{\texttt{#1}}
\expandafter\ifx\csname urlstyle\endcsname\relax
  \providecommand{\doi}[1]{doi: #1}\else
  \providecommand{\doi}{doi: \begingroup \urlstyle{rm}\Url}\fi

\bibitem[Balle \& Mohri(2012)Balle and Mohri]{Balle2012spectral}
Balle, B. and Mohri, M.
\newblock Spectral learning of general weighted automata via constrained matrix
  completion.
\newblock \emph{Advances in Neural Information Processing Systems (NIPS)}, pp.\
   2168--2176, 2012.

\bibitem[Borcea et~al.(2002)Borcea, Papanicolaou, Tsogka, and
  Berryman]{Borcea2002imaging}
Borcea, L., Papanicolaou, G., Tsogka, C., and Berryman, J.
\newblock Imaging and time reversal in random media.
\newblock \emph{Inverse Problems}, 18\penalty0 (5):\penalty0 1247, 2002.

\bibitem[Cai et~al.(2010)Cai, Candes, and Shen]{cai2010singular}
Cai, J.~F., Candes, E.~J., and Shen, Z.
\newblock A singular value thresholding algorithm for matrix completion.
\newblock \emph{SIAM Journal on Optimization}, 20\penalty0 (4):\penalty0
  1956--1982, 2010.

\bibitem[Candes \& Fernandez-Granda(2012)Candes and
  Fernandez-Granda]{CandesFernandez2012SRNoisy}
Candes, E.~J. and Fernandez-Granda, C.
\newblock Super-resolution from noisy data.
\newblock \emph{Arxiv 1211.0290}, November 2012.

\bibitem[Candes \& Fernandez-Granda(2013)Candes and
  Fernandez-Granda]{CandesFernandez2012SR}
Candes, E.~J. and Fernandez-Granda, C.
\newblock Towards a mathematical theory of super-resolution.
\newblock \emph{to appear in Communications on Pure and Applied Mathematics},
  2013.

\bibitem[Candes \& Recht(2009)Candes and Recht]{ExactMC09}
Candes, E.~J. and Recht, B.
\newblock Exact matrix completion via convex optimization.
\newblock \emph{Foundations of Computational Mathematics}, 9\penalty0
  (6):\penalty0 717--772, April 2009.

\bibitem[Candes et~al.(2006)Candes, Romberg, and Tao]{CandRomTao06}
Candes, E.~J., Romberg, J., and Tao, T.
\newblock Robust uncertainty principles: exact signal reconstruction from
  highly incomplete frequency information.
\newblock \emph{IEEE Transactions on Information Theory}, 52\penalty0
  (2):\penalty0 489--509, Feb. 2006.

\bibitem[Chen \& Chi(2013)Chen and Chi]{ChenChi2013TIT}
Chen, Y. and Chi, Y.
\newblock Robust spectral compressed sensing via structured matrix completion.
\newblock \emph{Arxiv: 1304.8126}, May 2013.

\bibitem[Chi et~al.(2011)Chi, Scharf, Pezeshki, and
  Calderbank]{Chi2011sensitivity}
Chi, Y., Scharf, L.L., Pezeshki, A., and Calderbank, A.R.
\newblock Sensitivity to basis mismatch in compressed sensing.
\newblock \emph{IEEE Transactions on Signal Processing}, 59\penalty0
  (5):\penalty0 2182--2195, May 2011.

\bibitem[Dragotti et~al.(2007)Dragotti, Vetterli, and Blu]{DragVetterliBlu2007}
Dragotti, P.~L., Vetterli, M., and Blu, T.
\newblock Sampling moments and reconstructing signals of finite rate of
  innovation: {S}hannon meets strang-fix.
\newblock \emph{IEEE Trans on Signal Processing}, 55\penalty0 (5):\penalty0
  1741 --1757, May 2007.

\bibitem[Duarte \& Baraniuk(2012)Duarte and Baraniuk]{duarte2012spectral}
Duarte, M.F. and Baraniuk, R.G.
\newblock Spectral compressive sensing.
\newblock \emph{Applied and Computational Harmonic Analysis}, 2012.

\bibitem[Fazel et~al.(2003)Fazel, Hindi, and Boyd]{Fazel2003Hankel}
Fazel, M., Hindi, H., and Boyd, S.~P.
\newblock Log-det heuristic for matrix rank minimization with applications to
  {H}ankel and {E}uclidean distance matrices.
\newblock \emph{American Control Conference}, 3:\penalty0 2156 -- 2162 vol.3,
  June 2003.

\bibitem[Fazel et~al.(2011)Fazel, Pong, Sun, and Tseng]{Fazel2011hankel}
Fazel, M., Pong, T.~K., Sun, D., and Tseng, P.
\newblock Hankel matrix rank minimization with applications in system
  identification and realization, 2011.

\bibitem[Gedalyahui et~al.(2011)Gedalyahui, Tur, and
  Eldar]{gedalyahu2011multichannel}
Gedalyahui, K., Tur, R., and Eldar, Y.~C.
\newblock Multichannel sampling of pulse streams at the rate of innovation.
\newblock \emph{IEEE Trans on Sig. Proc}, 59:\penalty0 1491--1504, 2011.

\bibitem[Gross(2011)]{Gross2011recovering}
Gross, D.
\newblock Recovering low-rank matrices from few coefficients in any basis.
\newblock \emph{IEEE Transactions on Information Theory}, 57\penalty0
  (3):\penalty0 1548--1566, March 2011.

\bibitem[Hua(1992)]{Hua1992}
Hua, Y.
\newblock Estimating two-dimensional frequencies by matrix enhancement and
  matrix pencil.
\newblock \emph{IEEE Trans on Sig. Proc.}, 40\penalty0 (9):\penalty0 2267
  --2280, Sep 1992.

\bibitem[Lustig et~al.(2007)Lustig, Donoho, and Pauly]{Lustig2007sparse}
Lustig, M., Donoho, D., and Pauly, J.~M.
\newblock Sparse {MRI}: The application of compressed sensing for rapid {MR}
  imaging.
\newblock \emph{Magnetic Resonance in Medicine}, 58\penalty0 (6):\penalty0
  1182--1195, 2007.

\bibitem[Markovsky(2008)]{Markovsky2008structured}
Markovsky, I.
\newblock Structured low-rank approximation and its applications.
\newblock \emph{Automatica}, 44\penalty0 (4):\penalty0 891--909, 2008.

\bibitem[Negahban \& Wainwright(2012)Negahban and
  Wainwright]{Negahban2012restricted}
Negahban, S. and Wainwright, M.J.
\newblock Restricted strong convexity and weighted matrix completion: Optimal
  bounds with noise.
\newblock \emph{The Journal of Machine Learning Research}, 98888:\penalty0
  1665--1697, May 2012.

\bibitem[Nion \& Sidiropoulos(2010)Nion and Sidiropoulos]{NionSid2010}
Nion, D. and Sidiropoulos, N.~D.
\newblock Tensor algebra and multidimensional harmonic retrieval in signal
  processing for {MIMO Radar}.
\newblock \emph{IEEE Transactions on Signal Processing}, 58\penalty0
  (11):\penalty0 5693 --5705, Nov. 2010.

\bibitem[Potter et~al.(2010)Potter, Ertin, Parker, and
  Cetin]{potter2010sparsity}
Potter, L.C., Ertin, E., Parker, J.T., and Cetin, M.
\newblock Sparsity and compressed sensing in radar imaging.
\newblock \emph{Proceedings of the IEEE}, 98\penalty0 (6):\penalty0 1006--1020,
  2010.

\bibitem[Roy \& Kailath(1989)Roy and Kailath]{RoyKailathESPIRIT1989}
Roy, R. and Kailath, T.
\newblock Esprit-estimation of signal parameters via rotational invariance
  techniques.
\newblock \emph{IEEE Transactions on Acoustics, Speech and Signal Processing},
  37\penalty0 (7):\penalty0 984 --995, Jul 1989.

\bibitem[Sankaranarayanan et~al.(2010)Sankaranarayanan, Turaga, Baraniuk, and
  Chellappa]{Sankaranarayanan2010compressive}
Sankaranarayanan, A., Turaga, P., Baraniuk, R., and Chellappa, R.
\newblock Compressive acquisition of dynamic scenes.
\newblock \emph{ECCV 2010}, pp.\  129--142, 2010.

\bibitem[Sayeed \& Aazhang(1999)Sayeed and Aazhang]{SayeedAazhang1999}
Sayeed, A.~M. and Aazhang, B.
\newblock Joint multipath-{D}oppler diversity in mobile wireless
  communications.
\newblock \emph{IEEE Transactions on Communications}, 47\penalty0 (1):\penalty0
  123 --132, Jan 1999.

\bibitem[Schermelleh et~al.(2010)Schermelleh, Heintzmann, and
  Leonhardt]{schermelleh2010guide}
Schermelleh, L., Heintzmann, R., and Leonhardt, H.
\newblock A guide to super-resolution fluorescence microscopy.
\newblock \emph{The Journal of cell biology}, 190\penalty0 (2):\penalty0
  165--175, 2010.

\bibitem[Shin et~al.(2012)Shin, Larson, Ohliger, Elad, Pauly, Vigneron, and
  Lustig]{LustigHankel2013}
Shin, P., Larson, P., Ohliger, M., Elad, M., Pauly, J., Vigneron, D., and
  Lustig, M.
\newblock Calibrationless parallel imaging reconstruction based on structured
  low-rank matrix completion.
\newblock \emph{submitted to Magnetic Resonance in Medicine}, 2012.

\bibitem[Tang et~al.(2012)Tang, Bhaskar, Shah, and
  Recht]{TangBhaskarShahRecht2012}
Tang, G., Bhaskar, B.~N., Shah, P., and Recht, B.
\newblock Compressed sensing off the grid.
\newblock \emph{Arxiv 1207.6053}, July 2012.

\end{thebibliography}
\bibliographystyle{icml2013}
\end{document}